\documentclass[review]{elsarticle}

\usepackage{graphicx}
\usepackage{amsfonts,amssymb,dsfont,color,bbm}
\usepackage{tabularx,booktabs,multirow}
\usepackage{mathtools}
\usepackage{natbib}
\usepackage{geometry}
\usepackage{fleqn}
\usepackage{graphicx}
\usepackage{txfonts}
\usepackage{babel}
\usepackage{subcaption}
\usepackage{hyperref}
\usepackage{microtype}      
\usepackage[vlined,ruled,norelsize,lined,boxed]{algorithm2e}
\usepackage{tikz}
\usetikzlibrary{arrows.meta}

\newtheorem{theorem}{Theorem}
\newtheorem{lemma}{Lemma}
\newdefinition{remark}{Remark}
\newdefinition{problem}{Problem}
\newdefinition{example}{Example}
\newdefinition{proposition}{Proposition}
\newdefinition{definition}{Definition}
\newdefinition{assumption}{Assumption}
\newproof{proof}{Proof}

\allowdisplaybreaks

\DeclareMathAlphabet\mathbfcal{OMS}{cmsy}{b}{n}
\newcommand\hide[1]{}

\newcommand\BLANK{\mathfrak{E}}

\newcommand*\PR{\mathds{P}}
\newcommand\PSP{\mathcal{P}}

\newcommand*\EXP{\mathds{E}}

\newcommand{\bs}{\boldsymbol}
\DeclareMathOperator*{\argmin}{arg\,min}

\DeclareMathOperator\SPAN{span}

\author[1]{Nima Akbarzadeh\corref{cor1}}
\ead{nima.akbarzadeh@mail.mcgill.ca}
\author[2]{Aditya Mahajan}
\ead{aditya.mahajan@mcgill.ca}

\cortext[cor1]{Nima Akbarzadeh}
\address[1]{Department of Electrical and Computer Engineering, McGill
	University, 3480 Rue University, Montréal, QC H3A\,0E9.}
\address[2]{Department of Electrical and Computer Engineering, McGill
	University, 3480 Rue University, Montréal, QC H3A\,0E9.}

\begin{document}
	
\title{Two families of indexable partially observable restless bandits and Whittle index computation\tnoteref{t1}}
\tnotetext[t1]{This research was funded in part by Fonds de Recherche du Quebec-Nature et technologies (FRQNT).}

\begin{abstract}
We consider the restless bandits with general finite state space under partial observability with two observational models: first, the state of each bandit is not observable at all, and second, the state of each bandit is observable when it is selected. Under the assumption that the models satisfy a restart property, we prove that both models are indexable. For the first model, we derive a closed-form expression for the Whittle index. For the second model, we propose an efficient algorithm to compute the Whittle index by exploiting the qualitative properties of the optimal policy. We present detailed numerical experiments for multiple instances of machine maintenance problem. The result indicates that the Whittle index policy outperforms myopic policy and can be close to optimal in different setups.
\end{abstract}


\maketitle

\section{Introduction}

Resource allocation and scheduling problems arise in various applications including telecommunication networks, sensor management, patient prioritization, and machine maintenance. Restless bandits is a widely-used solution framework for such models \cite{meshram2018whittle,guha2010approximation,kaza2019sequential,kaza2018restless,aalto2016whittle,larranaga2016dynamic,akbarzadeh2019dynamic,villar2015multi,abad2016near,glazebrook2005index,villar2016indexability,qian2016restless,borkar2017whittle,borkar2017opportunistic,avrachenkov2018whittle}.

Identifying the optimal policy in restless bandits suffers from the curse of dimensionality because the state space is exponential in the number of alternatives~\cite{papadimitriou1999complexity}. To circumvent the curse of dimensionality, Whittle proposed an index heuristic which has a linear complexity in the number of alternatives~\cite{whittle1988restless}. The resulting policy, called the Whittle index policy, operates as follows: assign an index (called the Whittle index) to each state of each arm (or alternative) and then, at each time, play the arms in states with \textcolor{black}{the highest} indices. 

The Whittle index policy is attractive for two reasons. First, it is a scalable heuristic because its complexity is linear in the number of arms. Second, although it is a heuristic, there are certain settings where it is optimal \cite{gittins1979bandit,weber1990index,lott2000optimality,liu2010indexability} and, in general, it performs close to optimal in many instances \cite{nino2007dynamic,ansell2003whittle,glazebrook2005index,glazebrook2006some,ayesta2010modeling,akbarzadeh2019restart}.

Nonetheless, there are two challenges in using the Whittle index policy. First, the Whittle index heuristic is applicable only when a technical condition known as indexability is satisfied. There is no general test for indexability, and the existing sufficient conditions are only applicable for specific models \cite{ansell2003whittle,glazebrook2005index,glazebrook2006some,archibald2009indexability,avrachenkov2013congestion,ayesta2010modeling,glazebrook2013monotone,yu2018deadline}. Second, while there are closed-form expressions to compute the Whittle index in some instances~\cite{glazebrook2005index,glazebrook2006some,glazebrook2013monotone,aalto2016whittle,larranaga2016dynamic,kaza2018restless,kaza2019sequential,akbarzadeh2019restart,akbarzadeh2022conditions}, in general, the Whittle index policy has to be computed numerically. For a subclass of restless bandits which satisfy an additional technical condition known as PCL (partial conservation law), the Whittle index can be computed using an algorithm called the adaptive greedy algorithm~\cite{nino2001restless,nino2007dynamic}. Recently, \cite{akbarzadeh2022conditions,gast2022computing} presented generalizations of adaptive greedy algorithm which are applicable to all indexable restless bandits.

We are interested in resource allocation and scheduling problems where the state of each arm is not fully-observed. Such \textit{partially observable} restless bandit models are conceptually and computationally more challenging. The sufficient conditions for indexability that are derived for fully-observed bandits~\cite{whittle1988restless,weber1990index,glazebrook2005index,glazebrook2006some,glazebrook2013monotone,akbarzadeh2022conditions,ruiz2020multi} are not directly applicable to the partially observable setting. The existing literature on partially observable restless bandits often restricts attention to models where each arm has two states \cite{guha2010approximation,kaza2019sequential,kaza2018restless,villar2016indexability,abad2016near,borkar2017whittle,meshram2018whittle,aalto2016whittle,mate2020collapsing}. In some cases, it is also assumed that the two states are positively correlated \cite{kaza2018restless,kaza2019sequential,aalto2016whittle}; in others, it is assumed that the state dynamics are independent of the chosen action~\cite{larranaga2016dynamic,dance2019optimal,liu2021index}. There are very few results for \textcolor{black}{general finite state space} models under partial observability \cite{qian2016restless,akbarzadeh2019dynamic,larranaga2016dynamic,dance2019optimal,liu2021index}, and, for such models, indexability is often verified numerically. In addition, there are very few algorithms to compute the Whittle index for such models.

Recently, alternative index-based heuristics for partially observable restless bandits~\cite{brown2020index} \textcolor{black}{has been proposed}, but we restrict to Whittle index policy in this paper.

The main contributions of our paper are as follows:
\begin{itemize}
	\item We investigate partially observable restless bandits with \textcolor{black}{general finite state spaces} and consider two observation models, which we call model A and model B. We show that both models are indexable.
	\item For model A, we provide a closed-form expression to compute the Whittle index. For model B, we provide a refinement of the adaptive greedy algorithm of \cite{akbarzadeh2022conditions} to efficiently compute the Whittle index.
	\item We present a detailed numerical study which illustrates that the Whittle index policy performs close to optimal for small scale systems and outperforms a commonly used heuristic (the myopic policy) for large-scale systems.
\end{itemize}

The organization of the paper is as follows. In Section~\ref{sec:model}, we formulate the restless bandit problem under partial observations for two different models. Then, we define a belief state by which the partially-observable problem can be converted into a fully-observable one. In Section~\ref{sec:restless}, we present a short overview of restless bandits. In Section~\ref{sec:sufficient}, we show the restless bandit problem is indexable for both models and present a general formula to compute the index. In Section~\ref{sec:computation}, we present a countable state representation of the belief state and use it to develop methods to compute the Whittle index. In Section \ref{sec:proof}, we present the proofs of the results. In Section~\ref{sec:numerical}, we present a detailed numerical study which compares the performance of Whittle index policy with two baseline policies. Finally, we conclude in Section~\ref{sec:conclusion}.

\subsection{Notations and Definitions}
We use uppercase letters to denote random variables; the corresponding lowercase letter to denote their realization and corresponding calligraphic letters to denote the set of realizations. Superscripts index arms and subscripts index time, e.g., $X^i_t$ denotes the state of arm~$i$ at time~$t$. The subscript $0{:}t$ denote the history of the variable from time~$0$ to $t$, e.g., $X^i_{0:t}$ denotes $(X_0, \ldots, X_t)$. Bold letters denote the vector of variable for all arms, e.g., ${\bs X}_t$ denotes $(X^1_t, \ldots, X^n_t)$. \textcolor{black}{Given a collection of real numbers $p^1, \ldots, p^n$, we use $\prod_{i=1}^{n} p^i$ to denote their product $p^1 \cdot p^2 \cdot \dots \cdot p^n$.  Given a collection of sets ${\cal X}^1, \dots, {\cal X}^n$, we use $\prod_{i=1}^{n} {\cal X}^i$ to denote their Cartesian product ${\cal X}^1 \times {\cal X}^2 \times \dots \times {\cal X}^n$. For a finite set ${\cal X}$, ${\cal P}({\cal X})$ denote the set of probability mass functions (PMFs) on ${\cal X}$.}

We use $\mathds{I}$ as the indicator function, $\mathds{E}$ as the expectation operator, $\PR$ as the probability function, $\mathds{R}$ as the set of real numbers, $\mathds{Z}$ as the set of integers and $\mathds{Z}_{\ge m}$ as the set of integers that \textcolor{black}{are not lower than $m$}.

Given ordered sets $\mathcal{X}$ and $\mathcal{Y}$, a function $f: \mathcal{X} \times \mathcal{Y} \to \mathds{R}$ is called submodular if for any $x_1, x_2 \in \mathcal{X}$ and $y_1, y_2 \in \mathcal{Y}$ such that $x_2 \geq x_1$ and $y_2 \geq y_1$, we have $f(x_1, y_2) - f(x_1, y_1) \geq f(x_2, y_2) - f(x_2, y_1)$. Furthermore, the transition probability matrix~$P$ is stochastic monotone if for any $x, y \in \mathcal{X}$ such that $x < y$, we have $\sum_{w \in \mathcal X_{\ge z}} P_{xw} \leq \sum_{w \in \mathcal X_{\ge z}} P_{yw}$ for any $z \in \mathcal{X}$.
For any function $f \colon \mathcal{Z} \to \mathbb{R}$, 
$\SPAN(f)$ denotes the span semi-norm of~$f$, i.e., 
\(
\SPAN(f) = \sup_{z \in \mathcal{Z}} f(z) - \inf_{z \in \mathcal{Z}} f(z).
\)

\section{Model and Problem Formulation} \label{sec:model}

\subsection{Restless Bandit Process with restart}
A discrete-time restless bandit process (or arm) is a controlled Markov process~$(\mathcal{X}, \{0, 1\}, \allowbreak \{\bar{P}(a)\}_{a \in \{0, 1\}}, c, \pi_0, \mathcal{Y})$ where $\mathcal{X}$ denotes the finite set of states; $\{0, 1\}$ denotes the action space where the action~$0$ is called the \textit{passive} action and the action~$1$ is the \textit{active} action; $\bar{P}(a)$, $a \in \{0, 1\}$, denotes the transition matrix when action~$a$ is chosen; $c: \mathcal{X} \times \{0, 1\} \to \mathbb{R}_{\geq 0}$ denotes the cost function; $\pi_0$ denotes the initial state distribution; $\mathcal{Y}$ denotes the finite set of observations. 

\textcolor{black}{
\begin{assumption}[Restart property]
	All rows of the transition matrix $\bar{P}(1)$ are identical.
\end{assumption}
For models which satisfy Assumption~$1$, we denote $\bar{P}(0)$ by $P$ and denote each (identical) row of $\bar{P}(1)$ by $Q$. The term \textit{restart property} is used following the terminology of \cite{akbarzadeh2019restart}, where $Q$ was a PMF on the state space (i.e., on taking active action, the state resets according to PMF~$Q$). Note \cite{akbarzadeh2019restart} considered fully observed models, while we are considering partially observable setups.}

An operator has to select $m < n$ arms at each time but does not observe the state of the arms. We consider two observation models.
\begin{itemize}
	\item \textbf{Model A}: In model A, the operator does not observe anything. We denote this by $Y_t = \BLANK$, where $\BLANK$ denotes a blank symbol.
	\item \textbf{Model B}: In model B, the operator observes the state of the arm after it has been reset, i.e., 
	\begin{equation} \label{eq:observation}
		Y_{t+1} = \begin{cases}
			\BLANK ~ & \text{ if } ~ A_t = 0 \\
			X_{t+1} ~ & \text{ if } ~ A_t = 1
		\end{cases},
	\end{equation}
\end{itemize}
For model A, $\mathcal{Y} = \{\BLANK\}$ and for model B, $\mathcal{Y} = {\cal X} \cup \{\BLANK\}$. 


\subsection{Partially-observable Restless Multi-armed Bandit Problem}
A partially-observable restless multi-armed bandit (PO-RMAB) problem is a collection of $n$ independent restless bandits~$(\mathcal{X}^i, \{0, 1\}, \allowbreak \{P^i(a)\}_{a \in \{0, 1\}}, c^i, \pi^i_0, {\mathcal Y}^i)$, $i \in {\cal N} \coloneqq \{1, \ldots, n\}$. 

Let $\bs {\mathcal X} \coloneqq \prod_{i \in {\cal N}} {\mathcal X}^i$, \textcolor{black}{${\mathbfcal A}(m) \coloneqq \biggl\{ (a^1, \ldots, a^n) \in \{0, 1\}^n : \sum_{i \in {\cal N}} a^i = m \biggr\}$}, and $\bs {\mathcal Y} \coloneqq \prod_{i \in {\cal N}} {\mathcal Y}^i$ denote the combined state, action, and observation spaces, respectively. Also, let $\bs X_t = (X^1_t, \dots X^n_t) \in \bs{\mathcal X}$, $\bs A_t = (A^1_t, \dots, A^n_t) \in {\mathbfcal A}(m)$, and $\bs Y_t = (Y^1_t, \dots Y^n_t) \in \bs{\mathcal Y}$ denote the combined states, actions taken, and observations made by the operator at time~$t \geq 0$. Due to the independent evolution of each arm, for each realization ${\boldsymbol x}_{0:t}$ of ${\boldsymbol X}_{0:t}$ and ${\boldsymbol a}_{0:t}$ of ${\boldsymbol A}_{0:t}$, we have
\begin{align*}	
	\PR( \bs{X}_{t+1} = \bs{x}_{t+1} | \bs{X}_{0:t} = \bs{x}_{0:t},\bs{A}_{0:t} = \bs{a}_{0:t}) & = \prod_{i \in {\cal N}} \PR( X^i_{t+1} = x^i_{t+1} | X^i_{t} = x^i_{t}, A^i_{t} = a^i_{t}) \\
	& = \prod_{i \in {\cal N}} P^i_{x^i_t, x^i_{t+1}}(a^i_{t}).
\end{align*}
Let ${\bs \pi}_0 = \prod_{i \in {\cal N}} \pi^i_0$ denote the initial state distribution of all arms.

When the system is in state~${\boldsymbol x}_{t}$ and action~${\boldsymbol a}_{t}$ is taken, the system incurs a cost 
\( {\bs c}({\bs x}_t, {\bs a}_t) \coloneqq \sum_{i \in {\cal N}} c^i(x^i_t, a^i_t). \)
The action at time~$t$ is chosen according to 
\begin{align} 
	\label{eq:policy-g}
	{\bs A}_t = {\bs g}_t({\bs Y}_{0:t-1}, {\bs A}_{0:t-1}), 
\end{align}
where ${\bs g}_t$ is a \textcolor{black}{history-dependent} policy at time~$t$. Let ${\bs g} = (g_1, g_2, \ldots)$ denote the policy for infinite time horizon and let ${\mathbfcal G}$ denote the family of all such policies.
Then, the performance of policy~${\bs g}$ is given by
\textcolor{black}{\begin{equation}
	J^{({\bs g})} \coloneqq (1-\beta) \EXP\biggl[ \sum_{t = 0}^{\infty} \beta^t \sum_{i
		\in {\cal N}} c^i(X^i_t, A^i_t) \big| X^i_0 \sim \pi^i_0 \biggr], \label{eqn:obj_func}
\end{equation}}
where $\beta \in (0, 1)$ denotes the discount factor.

Formally, the optimization problem of interest is as follows: 
\begin{problem} \label{prob:infinite}
	Given a discount factor $\beta \in (0,1)$, the total number $n$ of arms, the number $m$ to be selected, the system model~$\{(\mathcal{X}^i, \{0, 1\}, P^i(a), c^i, \mathcal{Y}^i)\}_{i \in {\cal N}}$ of each arm, and the observation model at the operator, choose a history-dependent policy ${\bs g} \in {\mathbfcal G}$ that minimizes $J^{({\bs g})}$ given by~\eqref{eqn:obj_func}.
\end{problem}

\subsubsection*{Some remarks}
\begin{enumerate}
  \item 
    Problem~\ref{prob:infinite} is a POMDP and the standard methodology to solve POMDPs is to convert them to a fully observable Markov decision process (MDP) by viewing the ``belief state'' as the information state of the system~\cite{astrom1965optimal}. \textcolor{black}{We present such a belief state representation in Sec.~\ref{subsec:belief} and point out its limitations in the context of restless bandits.}

  \item \textcolor{black}{
      In Problem~\ref{prob:infinite}, the objective $J^{(g)}$ depends on the initial state distribution $(\pi^i_0)_{i \in \mathcal {N}}$. This can give the impression that the optimal policy may depend on the initial distribution. It is well known in the MDP literature that there exist policies that are optimal for all initial distributions~\cite{puterman2014markov}. However, our results rely on translating the belief state representation of the POMDP into a countable state MDP formulation and such a transformation is valid only when the initial state distribution is of a specific form. See Sec.~\ref{sec:computation} for details. Our results do not depend on the specific choice of the initial distribution, as long as it satisfies Assumption~\ref{assump:reachable} specified in Sec.~\ref{sec:computation}.
    }
  \item \textcolor{black}{
      In Problem~\ref{prob:infinite}, we consider the \emph{normalized} expected discounted cost as an objective, where the discounted cost is multiplied by $(1-\beta)$. In the MDP literature, one typically considers unnormalized objectives. However, normalized objective is typically used constrained MDPs~\cite{altman1999constrained} and has also been used in some of the previous literature on restless bandits~\cite{akbarzadeh2022conditions}. Note that multiplying the objective by a constant does not change the optimal policy. The reason that we use a normalized expected discounted cost is that it simplifies the description of the adaptive greedy algorithms to compute the Whittle index presented in Sec.~\ref{sec:computation}. 
    }
\end{enumerate}

\subsection{Some Examples}
In this section, we present some examples corresponding to the model presented
above.

\begin{example}\label{ex:sensor-network}
	Consider a sensor network where ther are $n$ sensors, each observing an independent Markov processes. We assume that the state $\{S^i_t\}_{t \ge 1}$ of each Markov process is integer valued and evolves in an auto-regressive manner: $S^i_{t+1} = S^i_t + W^i_t$, where $\{W^i_t\}_{t \ge 1}$ are i.i.d.\ processes which are also independent across the sensors. An estimator can observe only $m$, where $m < n$, sensors at each time. If a sensor is observed, the state of the Markov process at that sensor is revealed to the estimator. If a sensor is not observed, the estimator gets no new information about its state and has to estimate the state based on previous observations. The objective is to determine a sensor scheduling policy to decide which sensors to observe at each time.
	
    \textcolor{black}{In this case, it can be shown that when the noise processes $\{W^i_t\}_{t \ge 1}$ have symmetric and unimodal distributions, the optimal estimation strategy is a \emph{Kalman-filter} like strategy, i.e., the optimal estimate $\hat S^i_t$ is $S^i_t$ when the Markov process~$i$ is transmitted, and is equal to the previous estimate~$\hat S^i_{t-1}$ when the Markov process~$i$ is not transmitted~\cite{lipsa2011remote}. Thus, the error process $E^i_t \coloneqq S^i_{t} - \hat S^i_t$ has a restart property~\cite{chakravorty2017fundamental}. An instance of such a sensor network problem was considered in~\cite{mahajan2017remote}.
    }
\end{example}

\begin{example}\label{ex:machine-maintenance}
	Consider a maintenance company monitoring $n$ machines which are deteriorating independently over time. Each machine has multiple deterioration states sorted from \textit{pristine} to \textit{ruined} levels. However, the state of the machine is not observed. There is a cost associated with running the machine and the cost is non-decreasing function of the state. If a machine is left un-monitored, then the state of the machine deteriorates and after a while, it is ruined. However, the state of the machine is not observed.
	
	Furthermore, it is assumed the company cannot observe the state of the machines unless it sends a service-person to visit the machine. Replacing the machine is relatively inexpensive, and when service-persons visit a machine, \textcolor{black}{they} simply replace it with a new one. Due to manufacturing mistakes, all the machines may not be in pristine state when installed. If the service-person can observe the state of the machine when installing a new one, the observation model is same as model B. Otherwise, it is model A. There are $m$, where $m < n$, service-persons. The objective is to determine a scheduling policy to decide which machines should be serviced at each time. An instance of such machine maintenance problem in the context of maintaining demand response devices was considered in~\cite{abad2016near}.
\end{example}


\begin{example}
	Consider the problem of resource constrained health intervention delivery, where a community health center is monitoring $n$ patients to check if they are adhering to the prescribed medication~\cite{mate2020collapsing}. Each patient has a binary state of ``Adhering'' or ``Not Adhering'', which is hidden. There are $m$, where $m < n$, health workers, and if an health worker visits a patient, the state of the patient is observed. Moreover, it is assumed that after the visit by a health worker, the patient goes into the ``Adhering'' state. The objective is to determine a policy to schedule the health workers to maximize the number of patients in the ``Adhering'' state.
\end{example}

\subsection{Belief State} \label{subsec:belief}

Using standard results from Markov decision theory, the partially observable restless bandit problem can be converted to a fully observable restless bandit problem with belief (or posterior distribution) as states. We present the details in this section. Let define the operator's belief $\Pi^{i}_t \in \PSP({\cal X}^i)$ on the state of arm~$i$ at time~$t$ as follows: for any, $x^i_t \in {\cal X}^i$, let
\(
\Pi^{i}_t(x^i_t) \coloneqq \PR(X^i_t = x^i_t \,|\, Y^i_{0:t-1}, A^i_{0:t-1}).
\)
Note that $\Pi^{i}_t$ is a distribution-valued random variable. Also, define $\bs \Pi_t \coloneqq (\Pi^{1}_t, \dots, \Pi^{n}_t)$.

Then, for arm~$i$, the evolution of the belief state is as follows: 
for model A, the belief update rule is
\textcolor{black}{\begin{equation} \label{eq:pi-evolve-a}
	\Pi^i_{t+1} = 
	\begin{cases}
		\Pi^i_t P, & \hbox{if } A^i_t = 0, \\
		Q,    & \hbox{if } A^i_t = 1,
	\end{cases}
\end{equation}}
and for model B, the belief update rule is
\textcolor{black}{\begin{equation} \label{eq:pi-evolve-b}
	\Pi^i_{t+1} = 
	\begin{cases}
		\Pi^i_t P, & \hbox{if } A^i_t = 0, \\
		\delta^i_{X^i_{t+1}} \text{ where } X^i_{t+1} \sim Q, & \hbox{if } A^i_t = 1
	\end{cases}
\end{equation}
where $\delta_{x}$ is the Dirac delta distribution over the discrete state space ${
\cal X}$ with the value of one only for state~$x$.}
The per-step cost function of the belief state $\Pi^i_t$ when action $A^i_t$ is taken is
\[ \bar{c}(\Pi^i_t, A^i_t) = \mathds{E}[c^i_t(X^{i}_t, A^i_t) | Y^i_{0:t-1}, A^i_{0:t-1}] = \sum_{x \in {\cal X}^i} \Pi^{i}_t(x) c^i(x, A^i_t). \]

Define the combined belief state $\Theta_t \in \PSP(\bs{\cal X})$ of the system as follows: for any $\bs x \in \bs{\cal X}$, 
\[
\Theta_t(\bs x) = \PR(\bs X_t = \bs x \,|\, \bs Y_{0:t-1}, \bs A_{0:t-1}).
\]
Note that $\Theta_t$ is a random variable that takes values in
$\PSP(\bs{\cal X})$. Using standard results in
POMDPs~\cite{astrom1965optimal}, we have the following.
\begin{proposition} \label{prop:pomdp}
	In Problem~\ref{prob:infinite}, $\Theta_t$ is a sufficient statistic for
	$(\bs Y_{0:t-1}, \bs A_{0:t-1})$. Therefore, there is no loss
	of optimality in restricting attention to decision policies of the form $\bs A_t = g^{\text{belief}}_t(\Theta_t)$. Furthermore, an optimal policy with this structure can be identified by solving an appropriate dynamic program.
\end{proposition}
Next, we present our first simplification for the structure of optimal decision policy as follows.
\begin{proposition} \label{prop:belief}
	For any $\bs x \in \bs{\cal X}$, we have 
	\begin{equation} \label{eq:indep}
		\Theta_t(\bs x) = \prod_{i \in {\cal N}} \Pi^{i}_t(x^i),
		\quad \text{a.s.}.
	\end{equation}
	Therefore, there is no loss of optimality in restricting attention to
	decision policies of the form $\bs A_t = g^{\text{simple}}_t(\bs \Pi_t)$.
	Furthermore, an optimal policy with this structure can be identified by
	solving an appropriate dynamic program.
\end{proposition}
\begin{proof}
	Eq.~\eqref{eq:indep} follows from the conditional independence of the arms,
	and the nature of the observation function. The structure of the optimal
	policies then follow immediately from Proposition~\ref{prop:pomdp}. 
\end{proof}

In Propositions~\ref{prop:pomdp} and \ref{prop:belief}, we do not present the dynamic programs because they suffer from the curse of dimensionality. In particular, obtaining the optimal policy for PO-RMAB is PSPACE-hard~\cite{papadimitriou1999complexity}. So, we focus on the Whittle index heuristics to solve the problem. 


\section{Whittle index policy solution concept} \label{sec:restless}

For the ease of notation, we will drop the superscript $i$ from all relative variables for the rest of this and the next sections.

Consider an arm~$(\mathcal{X}, \{0, 1\}, \allowbreak \{\bar{P}(a)\}_{a \in \{0, 1\}}, c, \pi_0, \mathcal{Y})$ with a modified per-step cost function 
\begin{equation} \label{eqn:modif_cost}
	{\bar c}_\lambda(\pi, a) \coloneqq {\bar c}(\pi, a) + \lambda a, 
	\quad \forall \pi \in \PSP({\mathcal X}), 
	\forall a \in \{0, 1\}, \lambda \in \mathds{R}.
\end{equation}
The modified cost function implies that there is a penalty of $\lambda$ for taking the active action. Given any time-homogeneous policy $g: \PSP({\mathcal X}) \to \{0, 1\}$, the modified performance of the policy is 
\begin{equation}
	J^{(g)}_{\lambda} := (1-\beta) \EXP\biggl[ \sum_{t = 0}^{\infty} \beta^t \bar{c}_\lambda(\Pi_t, g(\Pi_t)) \big| \Pi_0 \biggr]. \label{eqn:obj_func-modif}
\end{equation}

Subsequently, consider the following optimization problem. 
\begin{problem}\label{prob:info-RB}
	Given an arm~$({\cal X}, {\cal Y}, \{0, 1\}, \{\bar{P}(a)\}_{a \in \{0, 1\}}, c)$, the discount factor~$\beta \in (0,1)$ and the penalty $\lambda \in \mathds{R}$, choose a Markov policy $g: \PSP({\mathcal X}) \to \{0, 1\}$ to minimize $J^{(g)}_{\lambda}$ given by \eqref{eqn:obj_func-modif}.
\end{problem}

Problem~\ref{prob:info-RB} is a Markov decision process where one may use dynamic programming to obtain the optimal solution as follows.
\begin{proposition} \label{prop:inf_RB}
	\textcolor{black}{Consider the fixed point equation 
	\begin{equation} \label{eqn:fixed_point}
		V_\lambda(\pi) = \min_{a \in \{0, 1\}} H_\lambda(\pi, a) 
	\end{equation}
	where for Model A we have
	\begin{align*}
		H_\lambda(\pi, 0) = (1-\beta)\bar{c}(\pi, 0) + \beta V_\lambda(\pi P), H_\lambda(\pi, 1) = (1-\beta)\bar{c}(\pi, 1) + (1-\beta) \lambda + \beta V_\lambda(Q)
	\end{align*}
	and for Model B, we have
	\begin{align*}
		H_\lambda(\pi, 0) = (1-\beta)\bar{c}(\pi, 0) + \beta V_\lambda(\pi P), H_\lambda(\pi, 1) = (1-\beta)\bar{c}(\pi, 1) + \beta \sum_{x \in {\cal X}} Q_x V_\lambda(\delta_x).
	\end{align*}
	Then \eqref{eqn:fixed_point} has a unique fixed point $V^*_\lambda$, and the policy
	\begin{align*}
		g_\lambda(\pi) = \begin{cases}
			0, & \text{ if } H_\lambda(\pi, 0) < H_\lambda(\pi, 1) \\
			1, & \text{ otherwise }
		\end{cases}
	\end{align*}
	is optimal for Problem~\ref{prob:info-RB}.}
\end{proposition}
\begin{proof} 
	The result follows immediately from Markov decision theory~\cite{puterman2014markov}. 
\end{proof}

Finally, we present the following definitions.
\begin{definition}[Passive Set] \label{def:passive}
	Given penalty~$\lambda$, define the passive set~${\cal W}_\lambda$ as the set of states where passive action is optimal for the modified arm, i.e., \[{\cal W}_\lambda := \left\{ \pi \in \Pi: g_\lambda(\pi) = 0 \right\}.\]
\end{definition}

\begin{definition}[Indexability] \label{def:idxbl}
	An arm is indexable if ${\cal W}_\lambda$ is non-decreasing in $\lambda$, i.e., for any $\lambda_1, \lambda_2 \in \mathds{R}$,
	\textcolor{black}{\begin{equation*}
		\lambda_1 < \lambda_2 \implies {\cal W}_{\lambda_1} \subseteq {\cal W}_{\lambda_2}.
	\end{equation*}}
\end{definition}
A restless multi-armed bandit problem is indexable if all $n$ arms are indexable.

\begin{definition}[Whittle index] \label{def:Widx}
	The Whittle index of the state~$\pi$ of an arm is the smallest value of $\lambda$ for which state~$\pi$ is part of the passive set ${\cal W}_\lambda$, i.e.,
	\begin{equation*}
		w(\pi) = \inf \left\{ \lambda \in \mathds{R}: \pi \in {\cal W}_\lambda \right\}.
	\end{equation*}
\end{definition}
Equivalently, the Whittle index $w(\pi)$ is the smallest value of $\lambda$ for which the optimal policy is indifferent between the active action and passive action when the belief state of the arm is $\pi$. 

The Whittle index policy is as follows: \textit{At each time step, select $m$ arms which are in states with the highest indices}. The Whittle index policy is easy to implement and efficient to compute but it may not be optimal. As mentioned earlier, Whittle index is optimal in certain cases~\cite{gittins1979bandit,weber1990index,lott2000optimality,liu2010indexability} and performs close to optimal for many other cases~\cite{nino2007dynamic,ansell2003whittle,glazebrook2005index,glazebrook2006some,ayesta2010modeling,akbarzadeh2019restart}.

\section{Indexability and the corresponding Whittle index for models A and B} \label{sec:sufficient}

Given an arm, let $\Sigma$ denote the family of all stopping times \textcolor{black}{$\tau \geq 1$,} with respect to the natural filtration associated with $\{ \Pi_t \}_{t \geq 0}$. For any stopping time $\tau \in \Sigma$ and an initial belief state~$\pi \in \Pi$, define
\begin{align*} 
	L(\pi, \tau) & \coloneqq \EXP\bigg[ \sum_{t = 0}^{\tau-1} \beta^{t} {\bar c}(\Pi_t, 0) + \beta^{\tau} {\bar c}(\Pi_\tau, 1) \Bigm| \Pi_0 = \pi \bigg], \\
	B(\pi, \tau) & \coloneqq \EXP[ \beta^{\tau} | \Pi_0 = \pi ].
\end{align*}

\begin{theorem} \label{Thm:whittle_indxblty}
	The PO-RMAB for model A and B is indexable. In particular, each arm is indexable and the Whittle index is given by 
	\begin{equation*}
		w(\pi) = \inf \left\{ \lambda \in \mathbb{R} : G(\pi) < \Omega_\lambda \right\},
	\end{equation*}
	where 
	\begin{align}
		G(\pi) & \coloneqq (1-\beta) \inf_{\tau \in \Sigma} 
		\frac{L(\pi, \tau) - {\bar c}(\pi, 1)}{1 - B(\pi, \tau)}, \label{eqn:G_def} \\
		\Omega_\lambda & \coloneqq \lambda + \beta V^{\textsc{next}}_1 \label{eqn:W_def},
	\end{align}
	and $V^{\textsc{next}}_1 = V_\lambda(Q)$ for model A and $V^{\textsc{next}}_1 = \sum_{x \in {\cal X}} Q_{x} V_\lambda(\delta_x)$ for model B.
\end{theorem}
\begin{proof}
	
	\textcolor{black}{Recall that we assert that $V_\lambda(\pi)$ and $\Omega_\lambda$ are non-decreasing in $\lambda$ for any $\pi \in \Pi$. Hence, for any policy $g: {\cal P}({\cal X}) \to \{0, 1\}$
		\begin{equation*}
			V^{(g)}_\lambda(\pi) = (1 - \beta)\mathbb{E}\bigg[ \sum_{t = 0}^{\infty} \beta^t \bar{c}_\lambda(\Pi_t, A_t) \bigg| \Pi_t = \pi \bigg]
		\end{equation*}
	where $\bar{c}_\lambda(\pi, a) = c_\lambda(\pi, a) + \lambda a$ is non-decreasing in $\lambda$ for any $\pi \in {\cal P}({\cal X})$ and $a \in \{0, 1\}$. From Markov decision theory we know that $V_\lambda(\pi) = \inf_{g: {\cal P}({\cal X}) \to \{0, 1\}} V^{(g)}_\lambda(\pi)$. Since the infimum of non-decreasing functions is non-decreasing, $V_\lambda(\pi)$ is non-decreasing in $\lambda$ for any $\pi \in {\cal P}({\cal X})$. Consequently, $V^{\textsc{next}}_1$ is non-decreasing which implies $\Omega_\lambda$ is non-decreasing in $\lambda$.}
	
	\textcolor{black}{Given any stopping time $\tau \in \Sigma$, let $h_{\tau}$ denote a policy that takes the
		passive action up to and including time $\tau-1$, takes the active action at
		time~$\tau$, and follows the optimal policy from time~$\tau+1$ onwards. The
		performance $C_\lambda(\pi, \tau)$ of policy $h_\tau$ is given by
		\begin{align}
			C_\lambda(\pi, \tau) & = (1-\beta) \EXP^{h_\tau}\bigg[ 
			\sum_{t = 0}^{\infty} \beta^{t} c_\lambda(\pi_t, A_t) \Bigm| \Pi_0 = \pi 
			\bigg] = (1-\beta) L(\pi, \tau) + \EXP [ \beta^\tau \Omega_\lambda | \Pi_0 = \pi ] \notag \\
			& = (1-\beta) L(\pi, \tau) + B(\pi, \tau) \Omega_\lambda. \label{eqn:htau_perf}
			\end{align}
		We use $h_0$ to denote a policy that takes active action at time $0$ and follows the optimal policy from time $1$ onwards. The performance $C_\lambda(\pi, 0)$ of policy $h_0$ is given by
		\begin{equation} \label{eqn:htausetzero}
			C_\lambda(\pi, 0) = (1-\beta) c(\pi,1) + \Omega_\lambda.
		\end{equation}
	The next result generalizes \cite[Lemma 2]{akbarzadeh2019restart}.}
	\begin{lemma} \label{lem:characterization}
		\textcolor{black}{The following characterizations of the passive sets are equivalent
		to Def.~\ref{def:passive}.
		\begin{enumerate}
			\item ${\cal W}_\lambda = \left\{ \pi \in \Pi : H_\lambda(\pi, 0) < H_\lambda(\pi, 1) \right\}$.
			\item ${\cal W}_\lambda = \left\{ \pi \in \Pi : \exists \sigma \in \Sigma \text{ such that } C_\lambda(\pi, \sigma) < C_\lambda(\pi, 0) \right\}$.
			\item ${\cal W}_\lambda = \left\{ \pi \in \Pi : G(\pi) < \Omega_\lambda \right\}$.
		\end{enumerate}}
	\end{lemma}

\begin{proof}
	\textcolor{black}{Characterization~1) follows from the dynamic program given in Proposition~\ref{prop:inf_RB}.
	Characterization~2) follows from the fact that $C_\lambda(\pi,0) =
	H_\lambda(\pi,1)$ and for $\pi \in {\cal W}_\lambda$, $C_\lambda(\pi,\sigma) =
	H_\lambda(\pi,0)$, where $\sigma$ is the hitting time of the set $\PSP({\mathcal X})
	\setminus {\cal W}_\lambda$. Characterization~3) follows from characterization~2)
	and rearranging the terms using~\eqref{eqn:htau_perf}
	and~\eqref{eqn:htausetzero}.}
\end{proof}
	\textcolor{black}{Note that $G(\pi)$ does not depend on~$\lambda$ while we showed that
	$\Omega_\lambda$ is non-decreasing in~$\lambda$. Hence, ${\cal W}_\lambda = \left\{ \pi \in \Pi : G(\pi) < \Omega_\lambda \right\}$ is
	non-decreasing in $\lambda$ by the lemma.} Thus, arm $i$ is indexable. The expression for the
	Whittle index in the theorem follows from the definitions.
\end{proof}


\section{Whittle index computation} \label{sec:computation}
Computing the Whittle index using the belief state representation is intractable in general. Inspired by the approach taken in \cite{shuman2010measurement}, we introduce a new information state which is equivalent to the belief state.


\subsection{Countable Information state} \label{subsec:info_states}
For models A and B, define
\(
{\cal R}_A = \bigl\{ Q P^k : k \in \mathds{Z}_{\ge 0} \bigr\}, ~ {\cal R}_B = \bigl\{ \delta_s P^k : s \in {\cal X}, k \in \mathds{Z}_{\ge 0} \bigr\}.
\)
\begin{assumption} \label{assump:reachable}
	For model A, $\pi_0 \in {\cal R}_A$ and for model B, $\pi_0 \in {\cal R}_B$.
\end{assumption}

For model A, define a process $\{K_t\}_{t \geq 0}$ as follows. The initial state $k_0$ is such that $\pi_0 = Q P^{k_0}$ and for $t > 0$, $K_t$ is given by
\begin{equation} \label{eqn:ok_dyn-k}
	K_{t} = 
	\begin{cases}
		0, & \text{ if } A_{t-1} = 1 \\
		K_{t-1}+1, & \text{ if } A_{t-1} = 0.
	\end{cases}
\end{equation}

Similarly, for model B, define a process $\{S_t, K_t\}_{t \geq 0}$ as follows. The initial state $(s_0, k_0)$ is such that $\pi_0 = \delta_{s_0} P^{k_0}$ and for $t > 0$, $K_t$ evolves according to \eqref{eqn:ok_dyn-k} and $S_t$ evolves according to 
\begin{equation} \label{eqn:ok_dyn-s}
	S_{t} = 
	\begin{cases}
		X_{t-1} \text{ where } X_{t-1} \sim Q, & \text{ if } A_{t-1} = 1 \\
		S_{t-1}, & \text{ if } A_{t-1} = 0.
	\end{cases}
\end{equation}

Note that once the first observation has been taken in both models, $K_t$ denotes the time elapsed since the last observation of arm~$i$ and, in addition in model B, $S_t$ denotes the last observed states of arm~$i$. Let $\bs S_t \coloneqq (S^1_t, \dots S^n_t)$ and $\bs K_t \coloneqq (K^1_t, \dots K^n_t)$. The relation between the belief state $\Pi_t$ and variables $S_t$ and $K_t$ is characterized in the following lemma.

\begin{figure}[t!] 
	\centering
	\begin{tikzpicture}
		\draw[thick] (0, 0) node[anchor=north]{(1, 0, 0)}
		-- (2.5, 0) node[anchor=north]{(0, 1, 0)}
		-- (1.25, 2.2) node[anchor=south]{(0, 0, 1)}
		-- cycle;
		
		\draw[dashed,->] (0, 0) -- (.5, .5);
		\draw[dashed,->] (.5, .5) -- (1, .25);
		\draw[dashed,->] (1, .25) -- (1.5, .75);
		\draw[->] (1.5, .75) -- (2.5, 0);
		
		\draw[fill] (.5, .5) circle [radius = 0.025cm];
		\draw[fill] (1, .25) circle [radius = 0.025cm];
		\draw[fill] (1.5, .75) circle [radius = 0.025cm];
		
	\end{tikzpicture}
	\caption{Belief state dynamics for a 3-state arm~$i$ in the simplex~$\PSP(\{1, 2, 3\})$. Dashed arrows show a sample realizations of the belief state evolution under $A_t = 0$ for three time steps and the solid arrow shows a sample realization of the belief state evolution under $A_t = 1$.}
	\label{fig:jumps}
\end{figure}
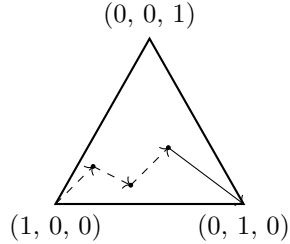

\begin{lemma} \label{lemma:reach_set}
	The following statements hold under Assumption~\ref{assump:reachable}:
	\begin{itemize}
		\item For model A, \textcolor{black}{for any $t$}, $\Pi_t \in {\cal R}_A$. In particular, $\Pi_t = Q P^{K_t}$.
		\item For model B, \textcolor{black}{for any $t$}, $\Pi_t \in {\cal R}_B$. In particular, $\Pi_t = \delta_{S_t} P^{K_t}$.
	\end{itemize}
\end{lemma}
\begin{proof}
	The results immediately follow from \eqref{eq:pi-evolve-a}-\eqref{eq:pi-evolve-b} and \eqref{eqn:ok_dyn-k}-\eqref{eqn:ok_dyn-s}. 
\end{proof}

For model A, the expected per-step cost at time~$t$ may be written as 
\begin{align} \label{eqn:bar_c-a}
	\bar{c}(K_t, A_t) \coloneqq \bar{c}(Q P^{K_t}, A_t) = \sum_{x \in {\cal X}} [Q P^{K_t}]_x c(x, A_t).
\end{align}

Similarly, for model B, the expected per-step cost at time~$t$ may be written as 
\begin{align} \label{eqn:bar_c-b}
	\bar{c}(S_t, K_t, A_t) \coloneqq \bar{c}(\delta_{S_t} P^{K_t}, A_t) = \sum_{x \in {\cal X}} [\delta_{S_t}, P^{K_t}]_x c(x, A_t).
\end{align}

\begin{proposition}
	In Problem~\ref{prob:infinite}, there is no loss of optimality in restricting attention to decision policies of the form $\bs A_t = g^{\text{info}}_t(\bs K_t)$ for model~A and of the form $\bs A_t = g^{\text{info}}_t(\bs S_t, \bs K_t)$ for model~B.
\end{proposition}
\begin{proof}
	This result immediately follows from Lemma~\ref{lemma:reach_set}, \eqref{eqn:bar_c-a} and \eqref{eqn:bar_c-b}. 
\end{proof}


\subsection{Threshold policies} \label{subsec:opt-pol}

We assume that the model satisfies the following condition.
\begin{assumption}\label{assump:cost}
	Let $c(x, a) = (1 - a) \phi(x) + a \rho(x)$ where $\phi: {\cal X} \to [0, \phi_{\max})$ and $\rho: {\cal X} \to [0, \rho_{\max})$ are non-decreasing functions in ${\cal X}$ and $c(x, a)$ is submodular in $(x, a)$.
\end{assumption}

Under Assumption~\ref{assump:cost}, we derive structural properties of the optimal policies for models A and B. Then, we show how $J^{(g)}_\lambda$ can be decomposed and computed.

In the following theorem, we show that the optimal policy for model A has a threshold structure and the optimal policy for model B has a threshold structure with respect to the second dimension of the information state. 
\begin{theorem} \label{Thm:opt-pol}
	Under Assumptions~\ref{assump:reachable} and \ref{assump:cost}, the following statements hold:
	\begin{enumerate}
		\item In model A, for any $\lambda \in \mathds{R}$, the optimal policy $g^A_\lambda(k)$ is a threshold policy, i.e., there exists a threshold~$\theta^A_\lambda \in \mathds{Z}_{ \geq -1 }$ such that
		\begin{align*}
			g^A_\lambda(k) = 
			\begin{cases}
				0, ~ k < \theta^A_\lambda \\
				1, ~ \text{otherwise}.
			\end{cases}
		\end{align*}
		Moreover, the threshold $\theta^A_\lambda$ is non-decreasing in
		$\lambda$.
		\item In model B, for any $\lambda \in \mathds{R}$, the optimal policy $g^B_\lambda(s, k)$ is a threshold policy with respect to $k$ for every $s \in {\cal X}$, i.e., there exists a threshold~$\theta^B_{s,\lambda} \in \mathds{Z}_{ \geq -1 }$ for each $s \in {\cal X}$ such that
		\begin{align*}
			g^B_\lambda(s, k) = 
			\begin{cases}
				0, ~ k < \theta^B_{s,\lambda} \\
				1, ~ \text{otherwise}.
			\end{cases}
		\end{align*}
		Moreover, for every $s \in {\cal X}$, the threshold
		$\theta^B_{s,\lambda}$ is non-decreasing in $\lambda$.
	\end{enumerate}
\end{theorem}
\textcolor{black}{See Section~\ref{sec:proof} for the proof.}

We use ${\bs \theta}^B$ to denote the vector $(\theta^B_{s})_{s \in {\cal X}}$. 

\subsection{Performance of threshold based policies} \label{subsec:perf-thresh}
We simplify the notation and denote the policy corresponding to thresholds~$\theta^A$ and ${\bs \theta}^B$ by simply $\theta^A$ and $\bs {\theta}^B$ instead of $g^{(\theta^A)}$ and $g^{(\bs {\theta}^B)}$.

\subsubsection{Model A}
Let $J^{(\theta^A)}_{\lambda}(k)$ be the total discounted cost incurred under
policy~$g^{(\theta^A)}$ with penalty~$\lambda$ when the initial state is $k$, i.e., 
\begin{align}
	J^{(\theta^A)}_{\lambda}(k) \coloneqq (1-\beta) \EXP \biggl[ \sum_{t = 0}^{\infty}
	\beta^t \bar{c}_\lambda(K_t, g^{(\theta^A)}(K_t)) \Bigm| K_0 = k \biggr] \eqqcolon D^{(\theta^A)}(k) + \lambda N^{(\theta^A)}(k), \label{eqn:Clambda}
\end{align}
where
\begin{align*}
	D^{(\theta^A)}(k) & \coloneqq (1-\beta) \EXP \biggl[ \sum_{t = 0}^{\infty} \beta^t c(K_t, g^{(\theta^A)}(K_t)) \Bigm| K_0 = k \biggr], \\
	N^{(\theta^A)}(k) & \coloneqq (1-\beta) \EXP \biggl[ \sum_{t = 0}^{\infty} \beta^t g^{(\theta^A)}(K_t) \Bigm| K_0 = k \biggr].
\end{align*}
$D^{({\theta^A})}(k)$ represents the expected total discounted cost while $N^{({\theta^A})}(k)$ represents the expected number of times active action is selected under policy~$g^{(\theta^{A})}$ starting from the initial information state~$k$.

We will show (see Theorem~\ref{Thm:whittle-a}) that the Whittle index for model A can be computed as a function of $D^{({\theta^A})}(k)$ and $N^{({\theta^A})}(k)$. First, we present a method to compute these two variables.
Let  
\begin{align*}
	L^{(\theta^A)}(k) & \coloneqq (1-\beta) \sum_{t = k}^{\theta^A-1} \beta^{t-k} \bar{c}(t, 0) + (1-\beta) \beta^{\theta^A-k} \bar{c}(\theta^A, 1) \\
	M^{(\theta^A)}(k) & \coloneqq (1-\beta) \beta^{\theta^A - k}
\end{align*}
where $L^{(\theta^A)}(k)$ and $M^{(\theta^A)}(k)$, respectively, denote the expected discounted cost and time starting from information state~$k$ until reaching information state~$\theta^A$ for the first time.
\begin{theorem} \label{Thm:DN-funcs-a}
	For any $k \in \mathds{Z}_{\geq 0}$, we have 
	\begin{align*}
		& D^{(\theta^A)}(k) = L^{(\theta^A)}(k) + \beta^{\theta^A-k+1} \dfrac{L^{(\theta^A)}(0)}{1 - \beta^{\theta^A+1}}, \\
		& N^{(\theta^A)}(k) = M^{(\theta^A)}(k) + \beta^{\theta^A-k+1} \dfrac{M^{(\theta^A)}(0)}{1 - \beta^{\theta^A+1}}.
	\end{align*}
\end{theorem}
\textcolor{black}{See Section~\ref{sec:proof} for the proof.}

\subsubsection{Model B}
Let $J^{({\bs \theta}^B)}_{\lambda}(s, k)$ be the total discounted cost incurred under
policy~$g^{({\bs \theta}^B)}$ with penalty~$\lambda$ when the initial information state is $(s, k)$, i.e.,
\begin{align}
	J^{({\bs \theta}^B)}_{\lambda}(s, k) & = (1-\beta) \EXP \biggl[ \sum_{t = 0}^{\infty} \beta^t \bar{c}_\lambda(S_t, K_t, g^{({\bs \theta}^B)}(S_t, K_t)) \Bigm| (S_0, K_0) = (s, k) \biggr] \nonumber \\
	& \eqqcolon D^{({\bs \theta}^B)}(s, k) + \lambda N^{({\bs \theta}^B)}(s, k) \label{eqn:C_decomposition},
\end{align}
where
\begin{align*}
	D^{({\bs \theta}^B)}(s, k) & \coloneqq (1-\beta) \EXP \biggl[ \sum_{t = 0}^{\infty} \beta^t \bar{c}(S_t, K_t, g^{({\bs \theta}^B)}(S_t, K_t)) \Bigm| (S_0, K_0) = (s, k) \biggr], \\
	N^{({\bs \theta}^B)}(s, k) & \coloneqq (1-\beta) \EXP \biggl[ \sum_{t = 0}^{\infty} \beta^t g^{({\bs \theta}^B)}(S_t, K_t) \Bigm| (S_0, K_0) = (s, k) \biggr].
\end{align*}
$D^{({\bs \theta}^B)}(s, k)$ and $N^{({\bs \theta}^B)}(s, k)$ have the same interpretations as the ones for model A. We will show (see Theorem~\ref{Thm:whittle-b}) that Whittle index for model B can be computed as a function of $D^{({\bs \theta^B})}(s, k)$ and $N^{({\bs \theta^B})}(s, k)$. 
But first let's define vector~$\boldsymbol{J}^{({\bs \theta}^B)}_{\lambda}(0) = (J^{({\bs \theta}^B)}_{\lambda}(1, 0)$, $\ldots$, $J^{({\bs \theta}^B)}_{\lambda}(|{\cal X}|, 0))$ and vectors~$\boldsymbol{D}^{({\bs \theta}^B)}(0)$ and $\boldsymbol{N}^{({\bs \theta}^B)}(0)$ in a similar manner. Then, from \eqref{eqn:C_decomposition}, $\boldsymbol{J}^{({\bs \theta}^B)}_{\lambda}(0) = \boldsymbol{D}^{({\bs \theta}^B)}(0) + \lambda \boldsymbol{N}^{({\bs \theta}^B)}(0)$. 
Let's also define 
\begin{align*}
	L^{({\bs \theta}^B)}(s, k) & \coloneqq (1-\beta) \sum_{t = k}^{\theta^B_s-1} \beta^{t-k} \bar{c}(s, t, 0) + (1-\beta) \beta^{\theta^B_s-k} \bar{c}(s, \theta^B_s, 1), \\
	M^{({\bs \theta}^B)}(s, k) & \coloneqq (1-\beta) \beta^{\theta^B_s - k}.
\end{align*}
Let $\boldsymbol{L}^{({\bs \theta}^B)}(0) = (L^{({\bs \theta}^B)}(1, 0), \ldots, L^{({\bs \theta}^B)}(|{\cal X}|, 0))$ and $\boldsymbol{M}^{({\bs \theta}^B)}(0) = (M^{({\bs \theta}^B)}(1, 0), \ldots, M^{({\bs \theta}^B)}(|{\cal X}|, 0))$.

\begin{theorem} \label{Thm:DN-funcs-b}
	For any $(s, k) \in{\cal X} \times \mathds{Z}_{\geq 0}$, we have 
	\begin{align*}
		& D^{({\bs \theta}^B)}(s, k) = L^{({\bs \theta}^B)}(s, k) + \beta^{\theta^B_s-k+1} \sum_{r \in{\cal X}} Q_r D^{({\bs \theta}^B)}(r, 0), \\
		& N^{({\bs \theta}^B)}(s, k) = M^{({\bs \theta}^B)}(s, k) + \beta^{\theta^B_s-k+1} \sum_{r \in{\cal X}} Q_r N^{({\bs \theta}^B)}(r, 0).
	\end{align*}
	
	Let $Z^{({\bs \theta}^B)}$ be a $|{\cal X}| \times |{\cal X}|$ matrix where $Z^{({\bs \theta}^B)}_{sr} = \beta^{\theta^B_s+1} Q_r$, for any $s, r \in {\cal X}$. Then, 
	\begin{align*}
		& \boldsymbol{D}^{({\bs \theta}^B)}(0) = (I - Z^{({\bs \theta}^B)})^{-1} \boldsymbol{L}^{({\bs \theta}^B)}(0), \\ & \boldsymbol{N}^{({\bs \theta}^B)}(0) = (I - Z^{({\bs \theta}^B)})^{-1} \boldsymbol{M}^{({\bs \theta}^B)}(0).
	\end{align*}
\end{theorem}
\textcolor{black}{See Section~\ref{sec:proof} for the proof.}

\subsection{Finite state approximation}
 
For computing Whittle index, we provide a finite state approximation of Proposition~\ref{prop:inf_RB} for models A and B. Essentially, we truncate the countable set of possible information state $K_t$ to a finite set and provide the approximation bound on the optimal value function for each of the models.
\begin{theorem}[Model A] \label{Thm:approx-infstat-a}
	Given $\ell \in \mathds{N}$, let $\mathds{N}_\ell := \{0, \ldots, \ell\}$ and $V_{\ell, \lambda} : \mathds{N}_\ell \to \mathds{R}$ be the unique fixed point of equation
	\begin{equation*}
		V_{\ell, \lambda}(k) = \min_{a \in \{0, 1\}} H_{\ell, \lambda}(k, a), ~ \hat{g}_{\ell, \lambda}(k) = \argmin_{a \in \{0, 1\}} H_{\ell, \lambda}(k, a) \label{eqn:vf-RB-approx}
	\end{equation*}
	where
	\begin{align*}
		H_{\ell, \lambda}(k, 0) & = (1-\beta)\bar{c}(k, 0) + \beta V_{\ell, \lambda}(\min\{k+1, \ell\}), \\ 
		H_{\ell, \lambda}(k, 1) & = (1-\beta)\bar{c}(k, 1) + (1-\beta) \lambda + \beta V_{\ell, \lambda}(0). \label{eqn:vf_aprx}
	\end{align*}
	We set $\hat{g}_{\ell, \lambda}(k) = 1$ if $H_{\ell, \lambda}(k, 0) = H_{\ell, \lambda}(k, 1)$. Then, we have the following: \\
	(i) For any $0 \leq k \leq \ell$, we have
	\begin{equation*}
		| V_\lambda(k) - V_{\ell, \lambda}(k) | \leq \dfrac{\beta^{\ell-k+1} \SPAN(c_{\lambda})}{1 - \beta}.
	\end{equation*}
	(ii) For all $k \in \mathds{Z}_{\geq 0}$, $\lim_{\ell \to \infty} V_{\ell, \lambda}(k) = V_{\lambda}(k)$. Moreover, let $\hat{g}^{*}_\lambda(\cdot)$ be any limit point of $\{\hat{g}_{\ell, \lambda}(\cdot)\}_{\ell \geq 1}$. Then, the policy $\hat{g}^{*}_\lambda(\cdot)$ is optimal for Problem~\ref{prob:info-RB}.
\end{theorem}
\textcolor{black}{See Section~\ref{sec:proof} for the proof.}

\begin{theorem}[Model B] \label{Thm:approx-infstat-b}
	Given $\ell \in \mathds{N}$, let $\mathds{N}_\ell := \{0, \ldots, \ell\}$ and $V_{\ell, \lambda} : {\cal X} \times \mathds{N}_\ell \to \mathds{R}$ be the unique fixed point of equation
	\begin{align*}
		V_{\ell, \lambda}(s, k) & = \min_{a \in \{0, 1\}} H_{\ell, \lambda}(s, k, a), ~ \hat{g}_{\ell, \lambda}(s, k) = \argmin_{a \in \{0, 1\}} H_{\ell, \lambda}(s, k, a)
	\end{align*}
	where
	\begin{align*}
		H_{\ell, \lambda}(s, k, 0) & = (1-\beta)\bar{c}(s, k, 0) + \beta V_{\ell, \lambda}(s, \min\{k+1, \ell\}), \\ 
		H_{\ell, \lambda}(s, k, 1) & = (1-\beta)\bar{c}(s, k, 1) + (1-\beta) \lambda + \beta \sum_{x' \in \tilde{\cal X}} Q_{x'} V_{\ell, \lambda}(x', 0).
	\end{align*}
	We set $\hat{g}_{\ell, \lambda}(s, k) = 1$ if $H_{\ell, \lambda}(s, k, 0) = H_{\ell, \lambda}(s, k, 1)$. Then, we have the following: \\
	(i) For any $0 \leq k \leq \ell$,
	\begin{equation*}
		| V_\lambda(s, k) - V_{\ell, \lambda}(s, k) | \leq \dfrac{\beta^{\ell-k+1} \SPAN(c_{\lambda})}{1 - \beta}, \forall s \in {\cal X}.
	\end{equation*}
	(ii) For all $(s, k) \in {\cal X} \times \mathds{Z}_{\ge 0}$, $\lim_{\ell \to \infty} V_{\ell, \lambda}(s, k) = V_{\lambda}(s, k)$. Let $\hat{g}^{*}_\lambda(\cdot, \cdot)$ be any limit point of $\{\hat{g}_{\ell, \lambda}(\cdot, \cdot)\}_{\ell \geq 1}$. Then, the policy $\hat{g}^{*}_\lambda(\cdot, \cdot)$ is optimal for Problem~\ref{prob:info-RB}.
\end{theorem}
\textcolor{black}{See Section~\ref{sec:proof} for the proof.}

Due to Theorems~\ref{Thm:approx-infstat-a} and \ref{Thm:approx-infstat-b}, we can restrict the countable part of the information state to a finite set, $\mathds{N}_\ell$.

\subsection{Computation of Whittle index} \label{subsec:whittle-idx}
Next, we derive a closed form expression to compute the Whittle index for model A and provide an efficient algorithm to compute the Whittle index for model B.
\subsubsection{Whittle index formula for model A. }
For model A, we obtain the Whittle index formula based on the two variables $D^{(\theta^A)}(\cdot)$ and $N^{(\theta^A)}(\cdot)$ as follows.
\begin{theorem} \label{Thm:whittle-a}
	\textcolor{black}{Let $\Lambda^A_k = \{k_0 \in \{0, 1, \ldots, (\ell+1)-1\} : N^{(k)}(k_0) \neq N^{(k+1)}(k_0)\}$. Then, under Assumption~\ref{assump:cost}, $\Lambda^A_k \neq \emptyset$, and the Whittle index of model A at information state $k \in \mathds{N}_\ell$ is}
	\begin{equation} \label{eqn:whittleindex-a}
		w^A(k) = \min_{k_0 \in \Lambda^A_k} \dfrac{D^{(k+1)}(k_0)-D^{(k)}(k_0)}{N^{(k)}(k_0)-N^{(k+1)}(k_0)}.
	\end{equation}
\end{theorem}
\begin{proof}
	Since model A is a restart model, the result follows from \cite[Lemma 4]{akbarzadeh2022conditions}. 
\end{proof}

Theorem~\ref{Thm:whittle-a} gives us a closed-form expression to approximately compute the Whittle index for model~$A$.

\subsubsection{Modified adaptive greedy algorithm for model B. }

Let $B = |{\cal X}|(\ell+1)$ and $B_D (\leq B)$ denote the number of distinct Whittle indices. Let $\Lambda^* = \{\lambda_0, \lambda_1, \ldots, \lambda_{B_D}\}$ where $\lambda_1 < \lambda_2 < \ldots < \lambda_{B_D}$ denote the sorted distinct Whittle indices with $\lambda_0 = -\infty$.
Let ${\cal W}_{b} := \{(s, k) \in {\cal X} \times \mathds{N}_\ell: w(s, k) \leq \lambda_b\}$. 
For any subset ${\cal S} \subseteq {\cal X}\times\mathds{N}_\ell$, define the policy $\bar{g}^{({\cal S})}: {\cal X}\times\mathds{N}_\ell \to \{0, 1\}$ as 
\begin{equation*}
	\bar{g}^{(\cal S)}(s, k) = 
	\begin{cases}
		0, ~ \text{ if } (s, k) \in {\cal S} \\
		1, ~ \text{ if } (s, k) \in ({\cal X}\times\mathds{N}_\ell)\backslash{\cal S}.
	\end{cases}
\end{equation*}

Given ${\cal W}_{b}$, define $\Phi_b = \{ (s, k) \in ({\cal X}\times\mathds{N}_\ell) \setminus {\cal W}_{b}: (s, \max\{0, k-1\}) \in {\cal W}_{b} \}$ and $\Gamma_{b+1} = {\cal W}_{b+1}\backslash{\cal W}_b$. Additionally, for any $b \in \{0, \ldots, B_D-1\}$, and all states $y \in \Phi_b$, define $h_b = \bar{g}^{({\cal W}_b)}$, $h_{b,y} = \bar{g}^{({\cal W}_b \cup \{y\})}$ and $\Lambda_{b, y} = \{(x, k) \in ({\cal X}\times\mathds{N}_\ell): N^{(h_b)}(x, k) \neq N^{({h}_{b,y})}(x, k) \}$. Then, for all $(x, k) \in \Lambda_{b, y}$, define
\begin{align} \label{eq:mu}
	\mu_{b,y}(x, k) = \dfrac{D^{({h}_{b, y})}(x, k) - D^{(h_b)}(x, k)}{N^{(h_b)}(x, k) - N^{({h}_{b, y})}(x, k)}. 
\end{align}
\begin{lemma} \label{lemma:WJ}
	For $d \in \{0, \ldots, B_D-1\}$, we have the following:
	\begin{enumerate}
		\item For all $y \in \Gamma_{b+1}$, we have $w(y) = \lambda_{b+1}$.
		\item For all $y \in \Phi_b$ and $\lambda \in (\lambda_{b}, \lambda_{b+1}]$, we have $J^{(h_{b,y})}_\lambda(x) \geq J^{(h_b)}_\lambda(x)$ for all $x \in {\cal X}$ with equality if and only if $y \in {\cal W}_{b+1}\backslash{\cal W}_{b}$ and $\lambda = \lambda_{b+1}$.
	\end{enumerate}
\end{lemma}
\begin{proof}
	The result follows from \cite[Lemma 3]{akbarzadeh2022conditions}. The only difference is that since we know from Theorem~\ref{Thm:opt-pol} that the optimal policy is a threshold policy with respect to the second dimension, we restrict to $y \in \Phi_b$. 
\end{proof}

\begin{theorem}\label{Thm:whittle-b}
	The following properties hold:
	\begin{enumerate}
		\item For any $y \in \Gamma_{b+1}$,
		the set $\Lambda_{b, y}$ is non-empty.
		\item For any $x \in \Lambda_{b, y}$, $\mu_{b,y}(x) \geq \lambda_{b+1}$ with equality if and only if $y \in \Gamma_{b+1}$.
	\end{enumerate}
\end{theorem}
\begin{proof}
	The result follows from  \cite[Theorem 2]{akbarzadeh2022conditions}. Similar to Lemma~\ref{lemma:WJ}, we consider $y \in \Phi_b$. 
\end{proof}
\begin{algorithm}[!t]
	\DontPrintSemicolon
	\SetKwInOut{Input}{input}
	\Input{RB~$(\mathcal{X}, \{0, 1\}, P, Q, c, \rho)$,
		discount factor $\beta$.} 
	Initialize $b = 0$, ${\cal W}_b = \emptyset$. \;
	\While{${\cal W}_b \neq {\cal X}\times\mathds{N}_\ell$}{
		Compute $\Lambda_{b, y}$ and $\mu_{b,y}(x)$ using~\eqref{eq:mu}, $\forall y \in \Phi_b$. \;
		Compute $\mu^*_{b,y} = \min_{x \in \Lambda_{b,y}} \mu_{b,y}(x)$, $\forall y \in \Phi_b$. \;
		Compute $\lambda_{b+1} = \min_{y \in \Phi_b} \mu^*_{b,y}$. \;
		Compute $\Gamma_{b+1} = \arg \min_{y \in \Phi_b} \mu^*_{b,y}$. \;
		Set $w(z) = \lambda_{b+1}$, $\forall z \in \Gamma_{b+1}$. \;
		Set ${\cal W}_{b+1} = {\cal W}_b \cup \Gamma_{b+1}$. \;
		Set $b = b+1$. \;
	}
	\caption{Computing Whittle index of all information states of model B}
	\label{Alg:Widy_comp}
\end{algorithm}
By Theorem \ref{Thm:whittle-b}, we can find the Whittle indices iteratively. This approach is summarized in Algorithm~\ref{Alg:Widy_comp}. For a computationally-efficient implementation using the Sherman-Morrison formula, see \cite[Algorithm 2]{akbarzadeh2022conditions}.


\section{Proof of Main Results} \label{sec:proof}

\subsection{Proof of Theorem~\ref{Thm:opt-pol}} \label{prf:opt-pol}
Let $\mu^1$ and $\mu^2$ be two probability mass functions on totally ordered set $\tilde{\cal X}$. Then we say $\mu^1$ \textit{stochastically dominates} $\mu^2$ if for all $x \in \tilde{\cal X}$, $\sum_{z \in \tilde{\mathcal X}_{\ge x}} \mu^1_z \geq \sum_{z \in \tilde{\mathcal X}_{\ge x}} \mu^2_z$. Given two $|\tilde{\cal X}| \times |\tilde{\cal X}|$ transition matrices $M$ and $N$, we say $M$ stochastically dominates $N$ if each row of $M$ stochastically dominates the corresponding $N$. A basic property of stochastic dominance is the following.

\begin{lemma} \label{lemma:sdom}
	If $M^1$ stochastically dominates $M^2$ and $c$ is an non-decreasing function defined on $\tilde{\cal X}$, then for all $x \in \tilde{\cal X}$, $\sum_{y \in \tilde{\cal X}} M^1_{xy} c(y) \geq \sum_{y \in \tilde{\cal X}} M^2_{xy} c(y)$.
\end{lemma}
\begin{proof}
	This follows from \cite[Lemma 4.7.2]{puterman2014markov}. 
\end{proof}

Consider a fully-observable restless bandit process $\{(\mathcal{\tilde X}, \{0, 1\}, \allowbreak \{\tilde P, \tilde Q\}, \tilde c, \tilde{\pi}_0)\}$ (note that ${\cal Y}$ is removed due to the observability assumption). According to \cite{akbarzadeh2022conditions}, we say a fully-observable restless bandit process is \emph{stochastic monotone} if it satisfies the following conditions.
\begin{enumerate}
	\item[\textup{(D1)}] $\tilde P$ and $\tilde Q$ are stochastic monotone transition matrices.
	\item[\textup{(D2)}] For any $z \in \tilde{\mathcal{X}}$, $\sum_{w \in \tilde{\mathcal X}_{\ge z}} [\tilde{P} - \tilde{Q}]_{xw}$ is non-decreasing in $x \in \tilde{\mathcal{X}}$.
	\item[\textup{(D3)}] For any $a \in \{ 0, 1 \}$, $\tilde c(x, a)$ is non-decreasing in $x$. 
	\item[\textup{(D4)}] $\tilde c(x, a)$ is submodular in $(x, a)$.
\end{enumerate}

The following is established in \cite[Lemma 5]{akbarzadeh2022conditions}.
\begin{proposition} \label{prop:threshold}
	The optimal policy of a stochastic monotone fully-observable restless bandit process is a \textit{threshold policy} denoted by $\tilde g$, which is a policy which takes passive action for states below a threshold denoted by $\tilde \theta$ and active action for the rest of the states, i.e., 
	\begin{align*}
		\tilde{g} = 
		\begin{cases}
			0, ~ x < \tilde \theta \\
			1, ~ \text{otherwise}
		\end{cases}.
	\end{align*}
\end{proposition}

\subsubsection{Proof of Theorem~\ref{Thm:opt-pol}, Part 1}
We show that each machine in model A is a stochastic monotone fully-observable restless bandit process.
Each condition of stochastic monotone fully-observable restless bandit process is presented and proven for model A below.	
\begin{enumerate}
	\item[(D1')] The transition probability matrix under passive action for model A based on the information states is $P^A_{xy} = \mathds{I}_{\{y = x+1\}}$ and the transition probability matrix under active action for model A is $Q^A_{xy} = \mathds{I}_{\{y = 0\}}$. Thus, $P^A$ and $Q^A$ are stochastic monotone matrices.
	\item[(D2')] Since $P^A$ is a stochastic monotone matrix and $Q^A$ has constant rows, $\sum_{r \geq z} [P^A - Q^A]_{sr}$ is non-decreasing in $s$ for any integer $z \geq 0$.
	\item[(D3')] As $P$ stochastically dominates the identity matrix, we infer from \cite[Theorem 1.1-b and Theorem 1.2-c]{keilson1977monotone}, that $QP^{\ell}$ stochastically dominates $QP^{k}$ for any $\ell > k \geq 0$. Additionally, $c_\lambda(x, a)$ is non-decreasing in $x$ for any $a \in \{0, 1\}$. By \eqref{eqn:bar_c-a} we have $\bar{c}_\lambda(k, a) = \sum_{x \in {\cal X}} [(QP)^{k}]_x c_\lambda(x, a)$. Therefore, by Lemma~\ref{lemma:sdom}, $\bar{c}_\lambda(k, a)$ is non-decreasing in $k$.
	\item[(D4')] As $c(x, a)$ is submodular in $(x, a)$ and as shown in (D3'), $QP^{\ell}$ stochastically dominates $QP^{k}$ for any $\ell > k \geq 0$. Therefore, by Lemma~\ref{lemma:sdom}, $\bar{c}_\lambda(k, 0) - \bar{c}_\lambda(k, 1) = \sum_{x \in {\cal X}} [(QP)^{k}]_x (c_\lambda(x, 0) - c_\lambda(x, 1))$ is non-decreasing in $(k, a)$. 
\end{enumerate}
Therefore, according to Proposition~\ref{prop:threshold}, the optimal policy of a fully-observable restless bandit process under model A is a threshold based policy.

Finally, since the optimal policy is threshold based, the passive set ${\cal
	W}_\lambda$ is given by $\{ k \in \mathbb{Z}_{\ge -1} : k < \theta^A_\lambda
\}$. As shown in Theorem~\ref{Thm:whittle_indxblty}, model~A is indexable.
Therefore, the passive set must be non-decreasing in $\lambda$, which implies
that the threshold $\theta^A_\lambda$ is non-decreasing in $\lambda$.

\subsubsection{Proof of Theorem~\ref{Thm:opt-pol}, Part 2}
We first characterize the behavior of value function and state-action value function for Model B.
\begin{lemma} \label{lemma:monotone}
	We have
	\begin{enumerate}
		\item[a.] $\bar{c}_\lambda(s, k, a)$ is non-decreasing in $k$ for any $s \in {\cal X}$ and $a \in \{0, 1\}$.
		\item[b.] Given a fixed $\lambda$, $V_\lambda(s, k)$ is non-decreasing in $k$ for any $s \in \mathcal{X}$.
		\item[c.] $\bar{c}_\lambda(s, k, a)$ is submodular in $(k, a)$, for any $s \in {\cal X}$.
		\item[d.] $H_\lambda(s, k, a)$ is submodular in $(k, a)$, for any $s \in {\cal X}$.
	\end{enumerate}
\end{lemma}
\begin{proof}
	The proof of each part is as follows.
	\begin{enumerate}
		\item[a.] By definition, we have
		\[ \bar{c}_\lambda(s, k, a) = \sum_{x \in {\cal X}} [\delta_s P^{k}](x) c(x, a) + \lambda a. \]
		Similar to the proof of (D3') in Proposition~\ref{prop:threshold}, for a given $s \in {\cal X}$ and $a \in \{0, 1\}$, $[\delta_s P^{k}](x)$ is non-decreasing in $k$ and $x$ and as $c(x, a)$ is non-decreasing in $x$, $\bar{c}_\lambda(s, k, a)$ is non-decreasing in $k$.
		\item[b.] 
		Let 
		\begin{align*}
			H^j_\lambda(s, k, 0) & := (1-\beta) \bar{c}(s, k, 0) + \beta V^j_\lambda(s, k+1), \\
			H^j_\lambda(s, k, 1) & := (1-\beta) \bar{c}(s, k, 1) + (1-\beta) \lambda + \beta \sum_{r} Q_r V^j_\lambda(r, 0), \\
			V^{j+1}_\lambda(s, k) & := \min_{a \in \{0, 1\}} \{H^j_\lambda(s, k, a)\},
		\end{align*}
		where $V^0_\lambda(\cdot, \cdot) = 0$ for all $(s, k) \in {\cal X} \times \mathds{Z}_{\geq 0}$.
		\\
		\textit{Claim:} $V^j_\lambda(s, k)$ is non-decreasing in $k$ for any $s \in {\cal X}$ and $j \geq 0$.
		\\
		We prove the claim by induction. By construction, $V^0_\lambda(s, k)$ is non-decreasing in $k$ for any $s \in {\cal X}$. This forms the basis of induction. Now assume that $V^{j}_\lambda(s, k)$ is non-decreasing in $k$ for any $s \in {\cal X}$ and some $j \geq 0$. Consider $\ell > k \geq 0$. Then, by induction hypothesis we have
		\begin{align*}
			H^j_\lambda(s, \ell, 0) & = (1-\beta) \bar{c}(s, \ell, 0) + \beta V^j_\lambda(s, \ell+1) \\
			& \geq (1-\beta) \bar{c}(s, k, 0) + \beta V^j_\lambda(s, k+1) = H^j_\lambda(s, k, 0), \\
			H^j_\lambda(s, \ell, 1) & = (1-\beta) \bar{c}(s, \ell, 1) + (1-\beta) \lambda + \beta \sum_{r} Q_r V^j_\lambda(r, 0) \\
			& \geq (1-\beta) \bar{c}(s, k, 1) + (1-\beta) \lambda + \beta \sum_{r} Q_r V^j_\lambda(r, 0) = H^j_\lambda(s, k, 1).
		\end{align*}
		Therefore,
		\begin{align*}
			V^{j+1}_\lambda(s, \ell) = \min_a \{H^j_\lambda(s, \ell, a)\} \geq \min_a \{H^j_\lambda(s, k, a)\} = V^{j+1}_\lambda(s, k).
		\end{align*}
		Thus, $V^{j+1}_\lambda(s, k)$ is non-decreasing in $k$ for any $s \in {\cal X}$. This completes the induction step.
		\( V_\lambda(s, k) = \lim_{j \to \infty} V^j_\lambda(s, k) \)
		and monotonicity is preserved under limits, the induction proof is complete.
		\item[c.] $c(x, a)$ is submodular in $(x, a)$. Also, note that $\delta_s P^{k}$ is the $s^{th}$ row of $P^{k}$. Thus, $\delta_s P^{k+1}$ stochastically dominates $\delta_s P^{k}$ and by Lemma~\ref{lemma:sdom} we have
		\[ \sum_{x \in {\cal X}} [\delta_s (P^{k+1} - P^{k})]_{x} (c(x, 0) - c(x, 1)) \geq 0. \]
		Therefore,
		\begin{align*}
			\sum_{x \in {\cal X}} [\delta_s (P^{k} - P^{k+1})]_{x} c(x, 1) \geq \sum_{x \in {\cal X}} [\delta_s (P^{k} - P^{k+1})]_{x} c(x, 0).
		\end{align*}
		Consequently,
		\begin{align*}
			\sum_{x \in {\cal X}} [\delta_s P^{k}]_{x} c(x, 1) - \sum_{x \in {\cal X}} [\delta_s P^{k}]_{x} c(x, 0) 
			\geq \sum_{x \in {\cal X}} [\delta_s P^{k+1}]_{x} c(x, 1) - \sum_{x \in {\cal X}} [\delta_s P^{k+1}]_{x} c(x, 0).
		\end{align*}
		Hence,
		\begin{align*}
			\bar{c}(s, k, 1) - \bar{c}(s, k, 0) \geq \bar{c}(s, k+1, 1) - \bar{c}(s, k+1, 0).
		\end{align*}
		\item[d.]
		As for any $s \in{\cal X}$, $V_\lambda(s, k)$ is non-decreasing in $k$, and $\bar{c}_\lambda(s, k, a)$ is submodular in $(k, a)$, for any $k \in \mathds{N}_\ell$ and $a \in \{0, 1\}$, we have
		\begin{align*}
			H_\lambda(s, k, 1) - H_\lambda(s, k, 0) & = (1-\beta)\bar{c}(s, k, 1) + (1-\beta)\lambda + \beta \sum_{r} Q_r V_\lambda(r, 0) \\
			& - (1-\beta)\bar{c}(s, k, 0) - \beta V_\lambda(s, k+1) \\
			& \geq (1-\beta)\bar{c}(s, k+1, 1) + (1-\beta)\lambda + \beta \sum_{r} Q_r V_\lambda(r, 0) \\
			& - (1-\beta)\bar{c}(s, k+1, 0) - \beta V_\lambda(s, k+2) \\
			& = H_\lambda(s, k+1, 1) - H_\lambda(s, k+1, 0). 
		\end{align*}
	\end{enumerate} 
\end{proof}

\begin{lemma} \label{lemma:sub-inc-pol}
	Suppose $f: \mathcal{X} \times \mathcal{Y} \to \mathds{R}$ is a submodular function and for each $x \in \mathcal{X}$, $min_{y \in \mathcal{Y}} f(x, y)$ exists. Then, $\max \{\argmin_{y \in \mathcal{Y}} f(x, y)\}$ is monotone non-decreasing in $x$.
\end{lemma}
\begin{proof}
	This result follows from \cite[Lemma 4.7.1]{puterman2014markov}. 
\end{proof}

Now, we conclude that as $H_\lambda(s, k, a)$ is submodular in $(k, a)$ for any $s \in {\cal X}$, then, based on Lemma~\ref{lemma:sub-inc-pol} and as only two actions is available, the optimal policy is a threshold policy specified in the theorem statement.

Finally, since the optimal policy is threshold based, the passive set ${\cal
	W}_\lambda$ is given by $\{ (s,k) \in {\cal X} \times \mathbb{Z}_{\ge -1} : k
< \theta^B_{s,\lambda} \}$. As shown in Theorem~\ref{Thm:whittle_indxblty},
model~B is indexable. Therefore, the passive set must be non-decreasing in
$\lambda$, which implies that, for every $s \in {\cal X}$, the threshold $\theta^B_{s,\lambda}$ is
non-decreasing in $\lambda$.

\subsection{Proof of Theorem~\ref{Thm:DN-funcs-a}} \label{prf:DN-funcs-a}
By the strong Markov property, we have 
\begin{align*}
	D^{(\theta^A)}(k) & = (1-\beta) \sum_{j = k}^{\theta^A} \beta^{t} \bar{c}(t, g(t)) + \beta^{\theta^A-k+1} D^{(\theta^A)}(0) = L^{(\theta^A)}(k) + \beta^{\theta^A-k+1} D^{(\theta^A)}(0), \\
	N^{(\theta^A)}(k) & = (1-\beta) \beta^{\theta^A-k} + \beta^{\theta^A-k+1} N^{(\theta^A)}(0) = M^{(\theta^A)}(k) + \beta^{\theta^A-k+1} N^{(\theta^A)}(0).
\end{align*}

If we set $k = 0$ in the above,
\begin{align*}
	D^{(\theta^A)}(0) & = \dfrac{L^{(\theta^A)}(0)}{1 - \beta^{\theta^A+1}} ~ \text{and} ~ 
	N^{(\theta^A)}(0) = \dfrac{M^{(\theta^A)}(0)}{1 - \beta^{\theta^A+1}}.
\end{align*}

\subsection{Proof of Theorem~\ref{Thm:DN-funcs-b}} \label{prf:DN-funcs-b}
By the strong Markov property, we have 
\begin{align*}
	D^{({\bs \theta}^B)}(s, k) & = (1-\beta) \sum_{j = k}^{\theta^B_s} \beta^{t} \bar{c}(s, t, g(s, t)) + \beta^{\theta^B_s-k+1} \sum_{r \in{\cal X}} Q_r D^{({\bs \theta}^B)}(r, 0) \\
	& = L^{({\bs \theta}^B)}(s, k) + \beta^{\theta^B_s-k+1} \sum_{r \in{\cal X}} Q_r D^{({\bs \theta}^B)}(r, 0), \\
	N^{({\bs \theta}^B)}(s, 0) & = (1-\beta) \beta^{\theta^B_s-k} + \beta^{\theta^B_s-k+1} \sum_{r \in{\cal X}} Q_r N^{({\bs \theta}^B)}(r, 0) \\
	& = M^{({\bs \theta}^B)}(s, k) + \beta^{\theta^B_s-k+1} \sum_{r \in{\cal X}} Q_r N^{({\bs \theta}^B)}(r, 0).
\end{align*}

If we set $k = 0$ in the above,
\begin{align*}
	D^{({\bs \theta}^B)}(s, 0) & = L^{({\bs \theta}^B)}(s, 0) + \beta^{\theta^B_s+1} \sum_{r \in{\cal X}} Q_r D^{({\bs \theta}^B)}(r, 0), \\
	N^{({\bs \theta}^B)}(s, 0) & = M^{({\bs \theta}^B)}(s, 0) + \beta^{\theta^B_s+1} \sum_{r \in{\cal X}} Q_r N^{({\bs \theta}^B)}(r, 0).
\end{align*}
which results in
\begin{align*}
	\boldsymbol{D}^{({\bs \theta}^B)}(0) & = \boldsymbol{L}^{({\bs \theta}^B)}(0) + Z^{({\bs \theta}^B)} \boldsymbol{D}^{({\bs \theta}^B)}(0), \\
	\boldsymbol{N}^{({\bs \theta}^B)}(0) & = \boldsymbol{M}^{({\bs \theta}^B)}(0) + Z^{({\bs \theta}^B)} \boldsymbol{N}^{({\bs \theta}^B)}(0)
\end{align*}
and hence, the statement is obtained by reformation of the terms inside the equations.

\subsection{Proof of Theorem~\ref{Thm:approx-infstat-a}}

(i): Starting from information state~$k \in \mathds{N}_\ell$, the cost incurred by $\hat{g}_{\ell, \lambda}(\cdot)$ is the same as $g^{A}_\lambda(\cdot)$ for information states~$\{k, \ldots, \ell\}$. The per-step cost incurred by $\hat{g}_{\ell, \lambda}(\cdot)$ differs from $g^{A}_\lambda(\cdot)$ for information states~$\{\ell+1, \ldots\}$ by at most $\SPAN(c_{\lambda})$. \\
(ii): The sequence of finite-state models described above is an \emph{augmentation type approximation sequence}. As a result, a limit point of $\hat{g}^{*}_\lambda$ exists and the final result holds by \cite[Proposition B.5, Theorem 4.6.3]{sennott2009stochastic}. 

\subsection{Proof of Theorem~\ref{Thm:approx-infstat-b}}

(i): Starting from information state~$(s, k)$, given any $s \in {\cal X}$ and $k \in \mathds{N}_\ell$, the cost incurred by $\hat{g}_{\ell, \lambda}(\cdot, \cdot)$ is the same as $g^B_{\lambda}(\cdot, \cdot)$ for information states~$\{(s, l)\}_{l = k}^{\ell}$. The per-step cost incurred by $\hat{g}_{\ell, \lambda}(\cdot, \cdot)$ differs from $g^B_{\lambda}(\cdot, \cdot)$ for later realized information states by at most $\Delta c_{\lambda}$. Thus, the bound holds. \\
(ii): The sequence of finite-state models described above is an \emph{augmentation type approximation sequence}. As a result, a limit point of $\hat{g}^{*}_\lambda$ exists and the final result holds \cite[Proposition B.5, Theorem 4.6.3]{sennott2009stochastic}. 


\section{Numerical Analysis} \label{sec:numerical}


In this section, we consider Example~\ref{ex:machine-maintenance} and compare the performance of the following policies:

\begin{itemize}
	\item[\textsc{opt}:] the optimal policy obtained using dynamic programming. As discussed earlier, the dynamic programming computation to obtain the optimal policy suffers from the curse of dimensionality. Therefore, the optimal policy can be computed only for small-scale models. 
	\item[\textsc{myp}:] myopic policy, which is a heuristic which sequentially selects $m$ machines as follows. Suppose $\zeta < m$ machines have been selected. Then select machine $\zeta + 1$ to be the machine which provides the smallest increase in the total per-step cost. The detailed description for model~B is shown in Alg.~\ref{Alg:MYP-b}.
	\item[\textsc{wip}:] whittle index heuristic, as described in this paper.
\end{itemize}


\begin{algorithm}[!h]
	\DontPrintSemicolon
	\SetKwInOut{Input}{input}
	\Input{RB~$(\mathcal{X}, \{0, 1\}, P, Q, c, \rho)$, discount factor $\beta$, $m$.} 
	Initialize $t = 0$. \;
	\While{$t \geq 0$}{
		Set $\zeta = 0$. \;
		\While{$\zeta \le m$}{
			Compute $i^*_\zeta \in \arg\min_{i \in \mathcal{Z}} \sum_{j \in \mathcal{Z}\setminus \{i\}} \bar{c}^j(S^j_t, K^j_t, 0) + \bar{c}^i(S^i_t, K^i_t, 1)$. \;
			Let $\mathcal{M} = \mathcal{M} \cup \{i^*_\zeta\}$, $\mathcal{Z} = \mathcal{Z} \setminus \{i^*_\zeta\}$. \;
			Set $\zeta = \zeta + 1$. \;
		}
		Service the machines with indices collected in $\mathcal{M}$. \;
		Update $K^i_t$ according to \eqref{eqn:ok_dyn-k} and $S^i_t$ according to \eqref{eqn:ok_dyn-s} for all $i \in {\cal N}$. \;
		Set $t = t+1$. \;
	}
	\caption{Myopic Heuristic (Model B)}
	\label{Alg:MYP-b}
\end{algorithm}

\subsection{Experiments and Results} \label{sebsec:setup}
We conduct numerical experiments for both models A and B, and vary the number~$n$ of machines, the number~$m$ of service-persons and the parameters associated with each machine.
There are three parameters associated with each machine: the deterioration probability matrix~$P^i$, the reset pmf~$Q^i$ and the per-step cost~$c^i(x, a)$. We assume the matrix~$P^i$ is chosen from a family of four types of structured transition matrices~${\cal P}_\gamma(p)$, $\gamma \in \{1, 2, 3, 4\}$ where $p$ is a parameter of the model. The details of all these models are presented \textcolor{black}{in \ref{app:markov-chain}}. We assume each element of $Q^i$ is sampled from Exp($1$), i.e., exponential distribution with the rate parameter of $1$, and then normalized such that the sum of all elements becomes $1$. Finally, we assume that the per-step cost is given by $c^i(x, 0) = (x-1)^2$ and $c^i(x, 1) = 0.5|\mathcal{X}^i|^2$.

In all experiments, the discount factor is $\beta = 0.99$. The performance of every policy is evaluated using Monte-Carlo simulation of length $1000$ averaged over $5000$ sample paths.

In Experiment 1, we consider a small scale problem where we can compute \textsc{opt} and we compare the performance of \textsc{wip} with it. However, in Experiment 2, we consider a large scale problem where we compare the performance of \textsc{wip} with \textsc{myp} as computing the optimal policy is highly time-consuming.

\textcolor{black}{The code for both experiments is available at \cite{code}.}

\subsubsection*{\textbf{Experiment 1)} Comparison of Whittle index with the optimal policy.} \label{subsec:exp1}
In this experiment, we compare the performance of \textsc{wip} with \textsc{opt}. We assume ${|\cal X|} = 4$, $(\ell+1) = 6$ and $n = 3$, $m = 1$ for both models A and B. In order to model heterogeneous machines, we consider the following. Let $(p_1, \ldots, p_n)$ denote $n$ equispaced points in the interval $[0.05, 0.95]$. Then we choose ${\cal P}_{\gamma}(p_i)$ as the transition matrix of machine~$i$. We denote the accumulated discounted cost of \textsc{wip} and \textsc{opt} by $J(\text{\textsc{wip}})$ and $J(\text{\textsc{opt}})$, respectively. In order to have a better perspective of the performances, we compute the relative performance of \textsc{wip} with respect to \textsc{opt} by computing
\begin{equation} \label{eqn:alpha}
	\alpha_{\textsc{opt}} = 100 \times \dfrac{J(\textsc{opt})}{J(\text{\textsc{wip}})}.
\end{equation}
The closer $\alpha$ is to $100$, the closer \textsc{wip} is to \textsc{opt}. The results of $\alpha_{\textsc{opt}}$ for all different combinations of parameter were $100$ which means the Whittle policy is as good as the optimal policy.

\subsubsection*{\textbf{Experiment 2)} Comparison of Whittle index with the myopic policy for structured models.} \label{subsec:exp3}
In this experiment, we increase the state space size to $|{\cal X}| = 20$ and we set $(\ell+1) = 40$,  we select $n$ from the set~$\{20, 40, 60\}$ and $m$ from the set $\{1, 5\}$. We denote the accumulated discounted cost of \textsc{myp} by $J(\text{\textsc{myp}})$. In order to have a better perspective of the performances, we compute the relative improvement of \textsc{wip} with respect to \textsc{myp} by computing
\begin{equation} \label{eqn:var_eps}
	\varepsilon_{\textsc{myp}} = 100 \times \dfrac{J(\text{\textsc{myp}})-J(\text{\textsc{wip}})}{J(\text{\textsc{myp}})}.
\end{equation}
Note that $\varepsilon_{\textsc{myp}} > 0$ means that \textsc{wip} performs better than \textsc{myp}. We generate structured transition matrices, similar to Experiment 1, and apply the same procedure to build heterogeneous machines. The results of $\varepsilon_{\textsc{myp}}$ for different choice of the parameters for models A and B are shown in Tables~\ref{tab:setup2-a} and \ref{tab:setup2-b}, respectively. 

\begin{table}[!t]
	\caption{$\varepsilon_{\textsc{myp}}$ for different choice of parameters of Model~A in Experiment~$2$.}\label{tab:setup2-a}
	\centering
	\begin{subtable}{0.5\linewidth}
		\centering
		\caption{Model A, $m = 1$}
		\begin{tabular}{cccccc}
			\toprule
			\multicolumn{2}{c}{\multirow{2}{*}{$\varepsilon_{\textsc{myp}}$}} & \multicolumn{4}{c}{$\gamma$} \\
			\cmidrule{3-6} 
			{} & {} & 1 & 2 & 3 & 4 \\
			\midrule
			\multirow{3}{*}{$n$} & $20$ & 1.99 & 2.54 & 2.24 & 7.44 \\ 
			{} & $40$ & 3.41 & 6.90 & 4.71 & 8.14 \\
			{} & $60$ & 2.97 & 6.19 & 2.80 & 6.70 \\
			\bottomrule
		\end{tabular}
	\end{subtable}%
	\begin{subtable}{0.5\linewidth}
		\centering
		\caption{Model A, $m = 5$}
		\begin{tabular}{cccccc}
			\toprule
			\multicolumn{2}{c}{\multirow{2}{*}{$\varepsilon_{\textsc{myp}}$}} & \multicolumn{4}{c}{$\gamma$} \\
			\cmidrule{3-6} 
			{} & {} & 1 & 2 & 3 & 4 \\
			\midrule
			\multirow{3}{*}{$n$} & $20$ & 0.21 & 0.26 & 0.19 & 0.97 \\ 
			{} & $40$ & 0.68 & 1.73 & 1.28 & 4.54 \\
			{} & $60$ & 1.36 & 2.35 & 2.32 & 6.41 \\
			\bottomrule
		\end{tabular}
	\end{subtable}
\end{table}
\begin{table}[!t]
	\caption{$\varepsilon_{\textsc{myp}}$ for different choice of parameters of Model~B in Experiment~$2$.}\label{tab:setup2-b}
	\centering
	\begin{subtable}{0.5\linewidth}
		\centering
		\caption{Model B, $m = 1$}
		\begin{tabular}{cccccc}
			\toprule
			\multicolumn{2}{c}{\multirow{2}{*}{$\varepsilon_{\textsc{myp}}$}} & \multicolumn{4}{c}{$\gamma$} \\
			\cmidrule{3-6}
			{} & {} & 1 & 2 & 3 & 4 \\
			\midrule
			\multirow{3}{*}{$n$} & $20$ & 7.67 & 11.17 & 12.12 & 9.39 \\ 
			{} & $40$ & 14.96 & 13.85 & 14.55 & 9.17 \\
			{} & $60$ & 15.02 & 12.12 & 13.39 & 6.63 \\
			\bottomrule
		\end{tabular}
	\end{subtable}%
	\begin{subtable}{0.5\linewidth}
		\centering
		\caption{Model B, $m = 5$}
		\begin{tabular}{cccccc}
			\toprule
			\multicolumn{2}{c}{\multirow{2}{*}{$\varepsilon_{\textsc{myp}}$}} & \multicolumn{4}{c}{$\gamma$} \\
			\cmidrule{3-6}
			{} & {} & 1 & 2 & 3 & 4 \\
			\midrule
			\multirow{3}{*}{$n$} & $20$ & 0.63 & 1.62 & 1.01 & 2.92 \\ 
			{} & $40$ & 2.92 & 3.14 & 3.21 & 6.57 \\
			{} & $60$ & 4.86 & 7.22 & 6.99 & 9.96 \\
			\bottomrule
		\end{tabular}
	\end{subtable}
\end{table}

\subsection{Discussion}
In Experiment 1 where \textsc{wip} is compared with \textsc{opt}, we observe $\alpha_\textsc{opt}$ is very close to $100$ for almost all experiments, implying that \textsc{wip} performs as well as \textsc{opt} for these experiments. $\alpha_\textsc{opt}$ in model B is less than model A as model B is more complex than model A for a given set of parameters and hence, the difference between the performance of the two polices is more than model A.

In Experiment 2 where \textsc{wip} is compared with \textsc{myp}, we observe $\varepsilon_\textsc{myp}$ ranges from $0.2 \%$ to $15 \%$. In a similar interpretation as Experiment 1, as model B is more complex than model A, $\varepsilon_\textsc{myp}$ for model B is higher than the ones model A given the same set of parameters. 

Furthermore, we observe that as $n$ increases, $\varepsilon_\textsc{myp}$ also increases overall. Also, as $m$ increases, $\varepsilon_\textsc{myp}$ decreases in general. This suggests that as $m$ increases, there is an overlap between the set of machines chosen according to \textsc{wip} and \textsc{myp}, and hence, the performance of \textsc{wip} and \textsc{myp} become close to each other.


\section{Conclusion} \label{sec:conclusion}
We investigated partially observable restless bandits. Unlike most of the existing literature which restricts attention to models with binary state space, we consider general state space models. We presented two observation models, which we call model A and model B, and showed that the partially observable restless bandits are indexable for both models. 

To compute the Whittle index, we work with a countable space representation rather than the belief state representation. We established certain qualitative properties of the auxiliary problem to compute the Whittle index. In particular, for both models we showed that the optimal policies of the auxiliary problem satisfy threshold properties. For model A, we used the threshold property to obtain a closed form expression to compute the Whittle index. For model B, we used the threshold policy to present a refinement of the adaptive greedy algorithm of \cite{akbarzadeh2022conditions} to compute the Whittle index.

Finally, we presented a detailed numerical study of a machine maintenance model. We observed that for small-scale models, the Whittle index policy is close-to-optimal and for large-scale models, the Whittle index policy outperforms the myopic policy baseline.

\appendix

\section{Structured Markov chains} \label{app:markov-chain}
Consider a Markov chain with $|\cal X|$ states. Then a family of structured stochastic monotone matrices which dominates the identity matrix is illustrated below.
\begin{enumerate}
	\item \textbf{Matrix ${\cal P}_{1}(p)$:} Let $q_1 = 1-p$ and $q_2 = 0$. Then,
	\begin{equation*}
		{\cal P}_{1}(p) = 
		\begin{bmatrix}
			p & q_1 & q_2 & 0   & 0   & 0   & 0 & \dots  & 0 \\
			0 & p   & q_1 & q_2 & 0   & 0   & 0 & \dots  & 0 \\
			0 & 0   & p   & q_1 & q_2 & 0   & 0 & \dots  & 0 \\
			0 & 0   & 0   & p   & q_1 & q_2 & 0 & \dots  & 0 \\
			\vdots & \vdots & \vdots & \vdots & \vdots & \vdots & \vdots & \vdots & \vdots \\
			0      & 0 & 0 & 0 & 0 & 0 & p & q_1  	& q_2 \\
			0      & 0 & 0 & 0 & 0 & 0 & 0 & p  	& q_1 + q_2 \\
			0      & 0 & 0 & 0 & 0 & 0 & 0 & \dots  & 1
		\end{bmatrix}.
	\end{equation*}
	\item \textbf{Matrix ${\cal P}_{2}(p)$:} Similar to ${\cal P}_{1}(p)$ with $q_1 = (1-p)/2$ and $q_2 = (1-p)/2$. 
	\item \textbf{Matrix ${\cal P}_{3}(p)$:} Similar to ${\cal P}_{1}(p)$ with $q_1 = 2(1-p)/3$ and $q_2 = (1-p)/3$.
	\item \textbf{Matrix ${\cal P}_{4}(p)$:} Let $q_i = (1-p)/({\cal X}-i)$. Then,
	\begin{align*}
		{\cal P}_{4}(p) = 
		\begin{bmatrix}
			p & q_1 & q_1 & \dots & q_1 & q_1 \\
			0 & p & q_2 & \dots & q_2 & q_2 \\
			\vdots & \vdots & \vdots & \vdots & \vdots & \vdots \\
			0 & 0 & 0 & \dots & p & q_{n-1} \\
			0 & 0 & 0 & \dots & 0 & 1
		\end{bmatrix}.
	\end{align*}
\end{enumerate}

\bibliographystyle{elsarticle-num}
\bibliography{mybibfile}

\end{document}